# Towards the continuum limit with quenched staggered quarks

Steven Gottlieb

Department of Physics, Indiana University, Bloomington, IN 47405, USA

We extend previous work on finite-size effects with dynamical staggered quarks to the quenched approximation. We again emphasize the large volume limit that is of interest for spectrum calculations which may hope to approach the experimental values. Relying on new calculations at $6/g^2 = 5.7$ and recent work with weaker couplings, we extrapolate to the continuum limit and find a nucleon to rho mass ratio in close agreement with the experimental value and the value obtained by extrapolations from calculations with Wilson quarks. Additional calculations that should be done to improve the reliability of the extrapolation are discussed.

## 1. MOTIVATION

A main goal of lattice QCD is to calculate the spectrum of light hadrons. Although there have been many spectrum calculations over the years, one persistent problem has been the large nucleon to rho mass ratio[1]. In any lattice calculation, there are certain limitations that may result in systematic bias in this ratio. For instance, the nucleon mass is more sensitive to box size than the rho, so for small volumes, $m_N/m_\rho$ is larger than in the infinite volume limit. Further, all calculations are done with quarks more massive than in nature and this also increases $m_N/m_\rho$. Recently, it has been shown using Wilson quarks in the quenched approximation that after extrapolating in quark mass, volume and lattice spacing that $m_N/m_\rho$ agrees quite well with the experimental value[2]. It would certainly be interesting to see if the same is true for staggered quarks.

We have been studying finite-size effects for quite some time using dynamical staggered quarks[3]. We realized that quenched results at $6/g^2 = 5.7$ [4], for which there is a comparable rho mass, apparently showed much larger finite-size effects. These quenched effects were also much larger than those recently seen at weaker coupling[5]. If we want to make finite volume corrections for weak coupling to avoid calculations with huge lattices, it is important to verify at stronger coupling that we understand how finite volume effects depend upon the quark mass (or $m_\pi/m_\rho$) and that they actually scale, *i.e.*, depend upon the physical box size.

In view of the above, we have begun a series of calculations with $6/g^2 = 5.7$. Combining our new results with those in the literature for 5.85, 5.95 [6] and 6.0 [5,7], we find that an extrapolation similar to that done with Wilson quarks yields a value of $m_N/m_\rho$ consistent with the experimental ratio.

## 2. PARAMETERS OF CALCULATION

Using $6/g^2 = 5.7$ we generated lattices of size $N_s^3 \times 48$ were generated using a combination of microcanonical and heat bath sweeps in the ratio of four to one. A total of 1000 sweeps were made between lattices on which the spectrum analysis was done. For each lattice, hadron propagators were calculated for quark masses $am_q = 0.16$, 0.08, 0.04, 0.02 and 0.01. The APE collaboration used the lowest four quark masses[4]. The heaviest one was added for better comparison with the Wilson quark calculations[2]. The time extent of these lattice is rather long. For the lighter masses, the hadron propagators cannot be accurately determined to distance 24; however, we use the large time size by calculating with six evenly spaced source planes on each lattice. Propagators were calculated for the $\pi$, $\pi_2$, $\rho$, $\rho_2$ and $N$, as well as other particles. Here we discuss only $\pi$, $\rho$ and $N$. (See Refs. 3 and 4 for more details of gauge fixing, sources, operators and other details of fitting.) At the time of the conference, we had analyzed 400 $N_s = 8$ lattices, 205 for $N_s = 12$ and 16 and 90 for $N_s = 20$. Since then, we have added 200 lattices with $N_s = 8$ using greater accuracy for the conjugate gradient algorithm and 60 lattices for $N_s = 20$ and 24. We have also carried out significant runs for $6/g^2 = 5.85$. In the interest of historical accuracy, the graphs presented here reflect none of the additional running, except for Fig. 3 where we could not resist including the new 5.85 result.

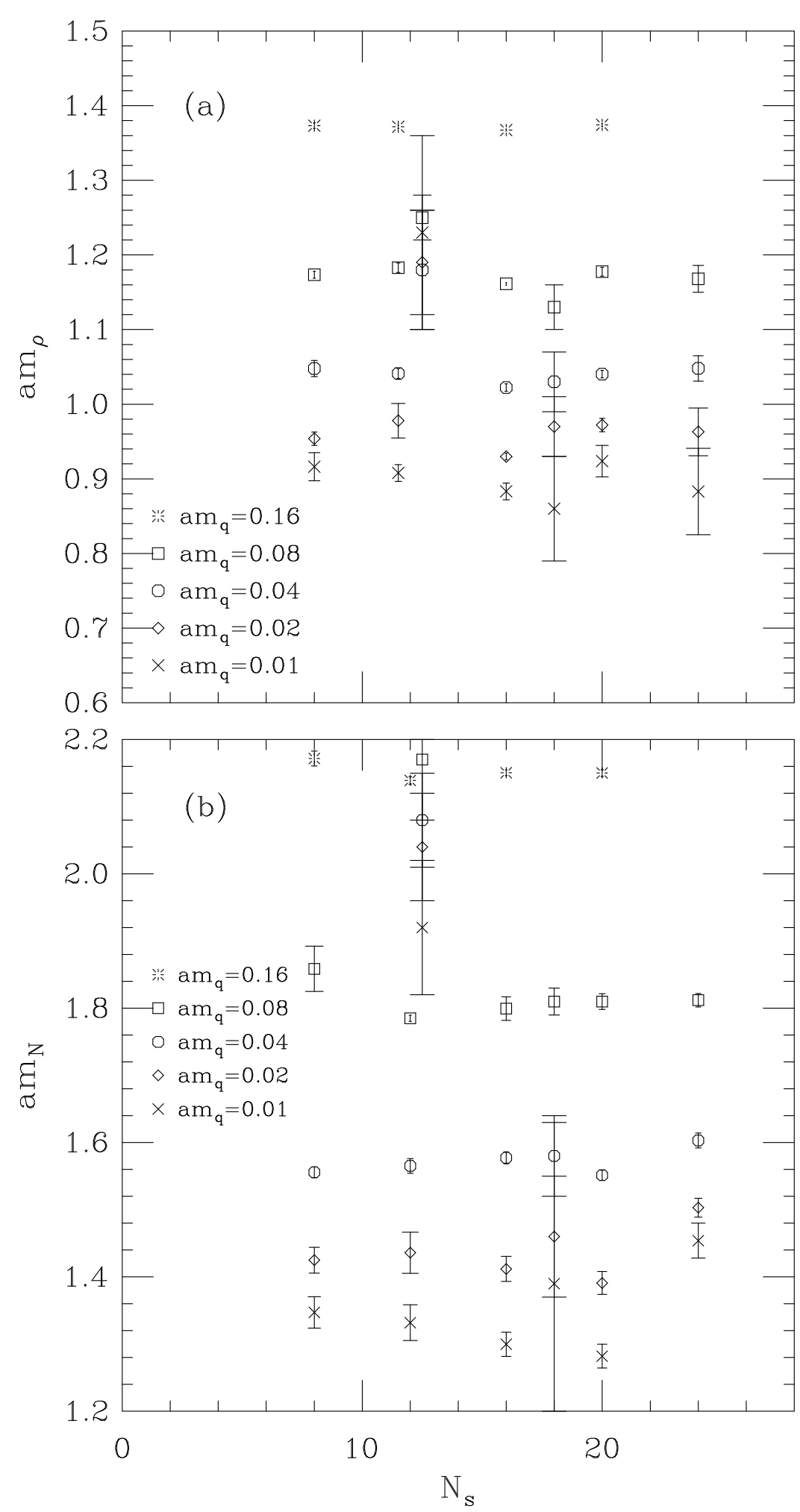

Figure 1. (a) Rho and (b) nucleon masses and lattice units as a function of $N_s$. Two sets of $N_s = 12$ results are displaced slightly left and right to avoid overlap.

## 3. RESULTS

In Fig. 1, we show our $\rho$ and $N$ masses and compare with the APE results for $N_s = 12$, 18 and 24 for which they analyzed 60, 18 and 50 lattices, respectively. For $N_s = 12$ the APE calculation uses point sources rather than smeared sources. Their apparently large finite-size effects are due to the source, not the box size. We find fairly small finite-size effects over the whole range of volumes shown here. For $N_s = 16$ the $\rho$ masses for the three lightest quark masses are somewhat low; however, all of the fits for this case have low confidence level. For the first half of the run, the masses were a little higher and the fits had higher confidence level. More running would be desirable in this case. Turning to the nucleon, for the two lightest masses, there apparently is some finite-size effect, as we find the masses dropping from $N_s = 8$ to 20. There is a large difference between the new $N_s = 20$ results and APE's $N_s = 24$ results for the three lightest quark masses. After the conference we analyzed 60 $N_s = 24$ lattices and find masses in good agreement with results for $N_s = 20$.

In Fig. 2, we show an Edinburgh plot that compares the Wilson results at this coupling to the staggered results. We find that for our largest volume the results, plotted in this way, are quite similar. For $N_s = 16$, there appear to be larger finite-size effects, but we emphasize how poor the $\rho$ fits were in that case. Our calculations extend somewhat further toward the chiral limit than for Wilson quarks.

With the masses in hand, we can attempt to extrapolate in quark mass, volume and lattice spacing. We found the finite-size effects to be small, so we just use our $N_s = 20$ results for the quark mass extrapolations. Linear fits to the $\rho$ and nucleon masses for all five quark masses are not good fits. Quadratic fits are acceptable, as are linear fits to the lightest four quark masses. We show in Fig. 3 $m_N/m_\rho$ at zero quark mass based upon our fits for $m_N$ and $m_\rho$. The figure also takes results from the literature from the HEMCGC group for $6/g^2 = 5.85$ and 5.95. For those two couplings there are only two quark masses available. We use an octagon for these linear fits. For $6/g^2 = 6.0$ we have combined the results of Aoki *et al.* [5] and Kim and Sinclair [7] for $0.005 \leq am_q \leq 0.02$ to do the chiral extrapolation. Using the four sets of linear fits, we fit $m_N/m_\rho$ to a linear function of lattice spacing and find the extrapolated value at the left of the figure. Our extrapolation compares quite favorably with the experimental result of 1.22. This graph also shows two diamonds that are chiral extrapolations of $\rho$ and $N$ masses based on quadratic fits. Also, we have added a point, the "fancy square," from new running at 5.85 done after the conference. This point is not used in making the extrapolation in lattice spacing. It's error is much smaller than the HEMCGC value at 5.85, and it supports the extrapolation shown at the conference.

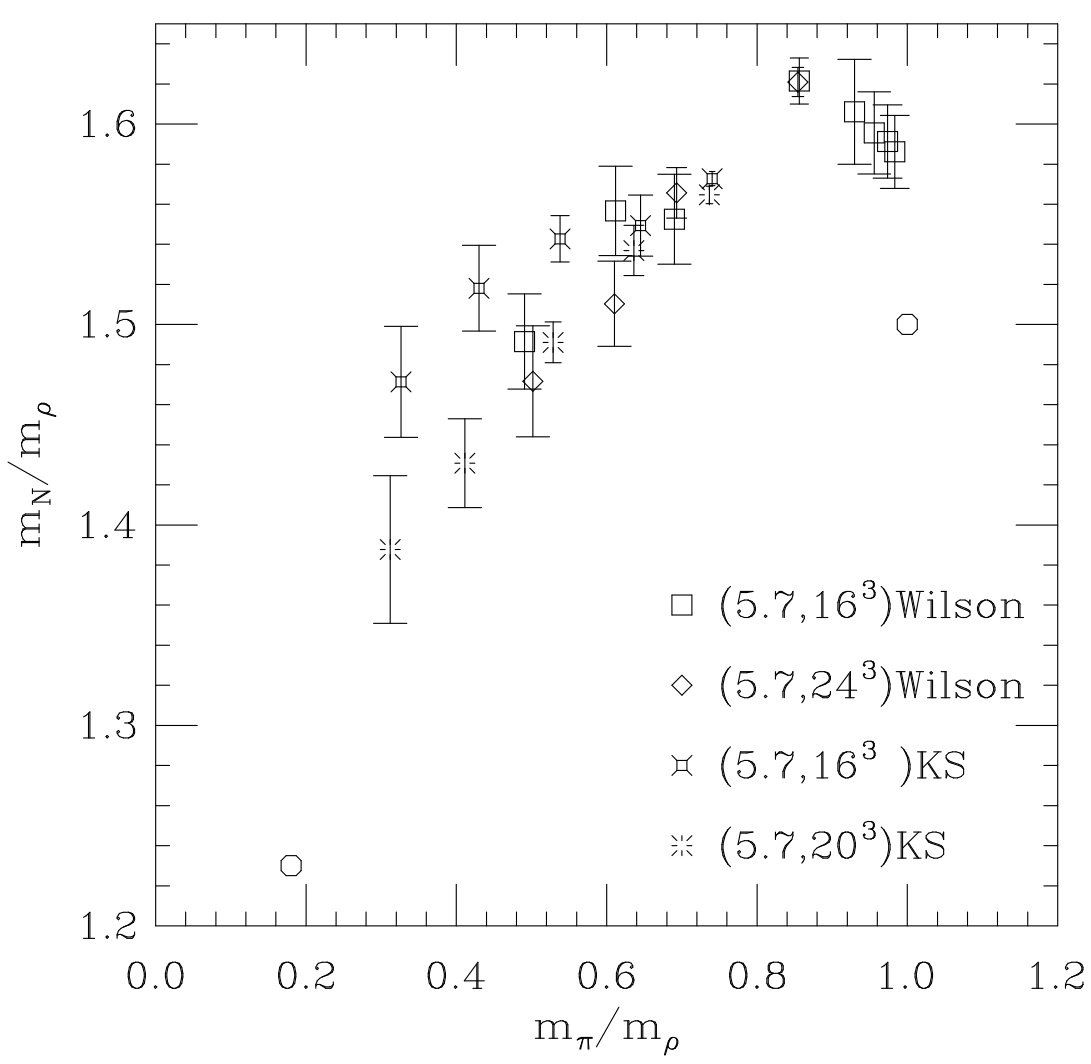

Figure 2. Edinburgh plot comparing Wilson and staggered quarks at $6/g^2 = 5.7$.

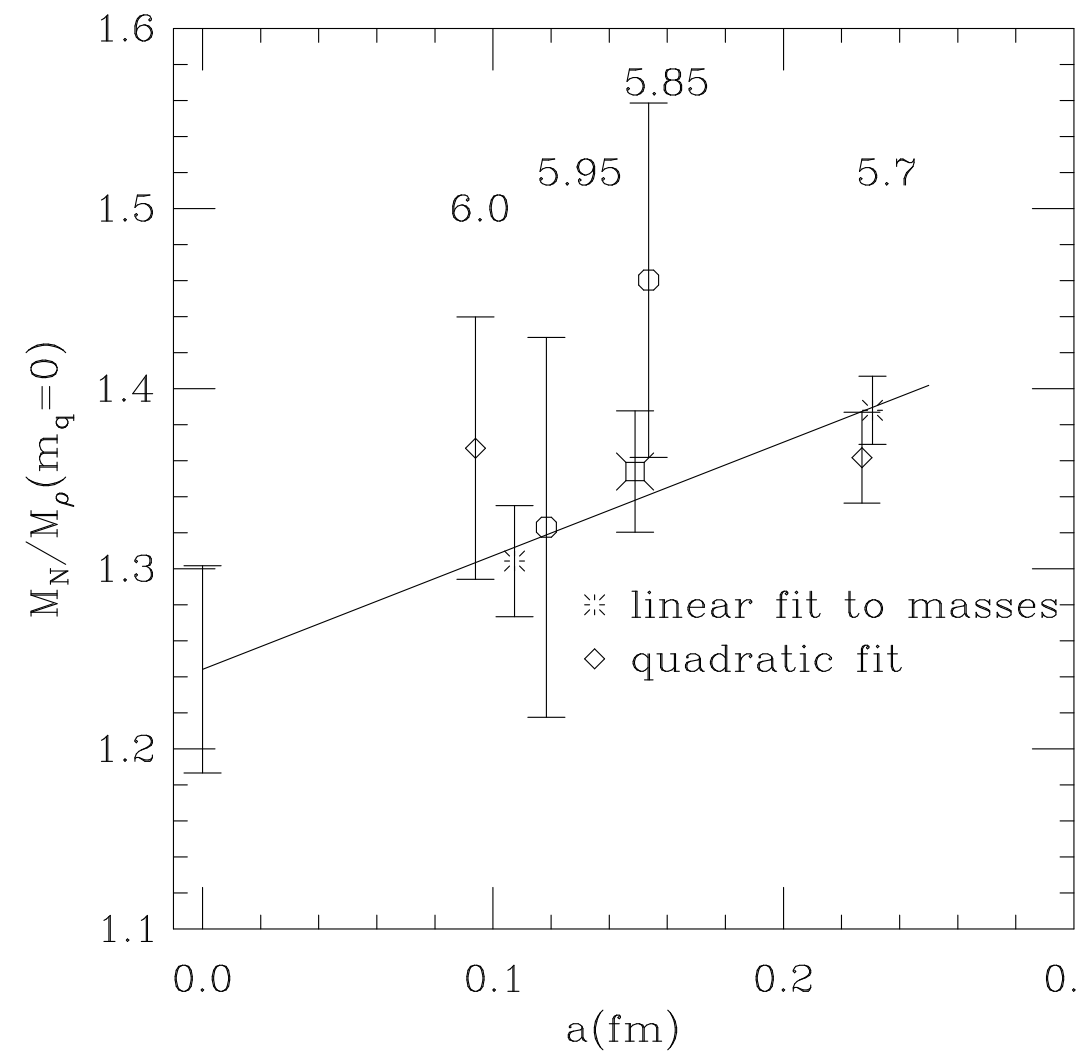

Figure 3. $m_N/m_\rho$ extrapolated to the zero quark mass limit as a function of lattice spacing.

In conclusion, the current calculation forms a firm basis for extrapolation in lattice spacing. Combining the current work with results in the literature, the extrapolation in lattice spacing of $m_N/m_\rho$ appears comparable to what has been seen with Wilson quarks. We have already begun to follow up on the current calculation by studying $6/g^2 = 5.85$ (albeit not yet on such large physical volumes as at 5.7), and have plans to study 6.15.

This work was supported by the U.S. Department of Energy Grant FG02-91ER 40661. All the calculations were done on the Indiana University Paragon parallel computer and we are grateful to IU Computing Services for their support, and to members of the MILC collaboration with whom we have done many prior calculations.